# New silicon photonics integration platform enabled by novel micron-scale bends

**Matteo Cherchi,**[*] **Sami Ylinen, Mikko Harjanne, Markku Kapulainen, and Timo Aalto**

*VTT Technical Research Centre of Finland, Espoo, 02040, Finland*

[*]Corresponding author: matteo.cherchi@vtt.fi

Abstract. Even though submicron silicon waveguides have been proposed for dense integration of photonic devices, to date the lightwave circuits on the market mainly rely on waveguides with micron-scale core dimensions. These larger waveguides feature easier fabrication, higher reliability and better interfacing to optical fibres. Single-mode operation with large core dimensions is obtained with low lateral refractive index contrast. Hence, the main limitation in increasing the level of integration and in reducing the cost of micron-scale waveguide circuits is their mm- to cm-scale minimum bending radius. Fortunately, single-mode rib waveguides with a micron-scale silicon core can be locally transformed into multi-mode strip waveguides that have very high lateral index contrast. Here we show how Euler spiral bends realized with these waveguides can have bending radii below 10 μm and losses below 0.02 dB/90° for the fundamental mode, paving way for a novel densely integrated platform based on micron-scale waveguides.

One of the main limitations in reducing the footprint and cost of photonic integrated circuits is the minimum bending radius of the waveguides. It depends strongly on the size and lateral refractive index contrast of the waveguide, and it can be chosen to reflect, for example, 1% bending loss per 90°. High index contrast (HIC) waveguides with submicron mode field dimensions can use µm-scale bending radii, whereas single-mode waveguides with mode field dimensions close to those of single-mode fibers require cm-scale radii.

An almost untouchable sacred rule in the design of integrated photonic circuits reads that, when reaching for small footprint and low propagation losses, any bent waveguide must be single-mode [1-3]. This is to avoid unwanted coupling to higher order modes (HOMs) and subsequent detrimental mode beating and power radiation in the bend. The densest integration and smallest bending radii are achieved with submicron HIC waveguides, and circuits based on them are being actively developed [1,3]. However, they pose many challenges and barriers to widespread use. Their strong polarization dependence requires either polarization diversity approaches [4] or a fully polarization maintaining system. They need unconventional waveguide crossings [5] to reduce losses, back reflection, and cross-talk. Resolving submicron features in scalable production requires expensive state-of-the-art fabrication tools [6] and involves high sensitivity to nanometre-scale fabrication errors [7]. And finally, coupling light between ultra-small HIC waveguides and standard single-mode fibers is a major challenge [8], especially when polarisation independence, broadband operation or low losses are required.

Using micron-scale single-mode waveguides now leads to bends and circuits with mm or even cm-scale dimensions, but also offers many benefits, such as small birefringence, low-loss crossing, efficient and broadband coupling to standard single-mode fibers, robustness to fabrication errors, and less expensive fabrication tools. Reducing the minimum bending radius of these larger waveguides would enable to add to their benefits also small size and cost, while circumventing the drawbacks of the submicron waveguides. Here we present a solution to achieve this goal by using HIC multi-mode waveguide bends that have bending radii below 10 µm and bending losses below 0.02 dB/90°. The experimentally confirmed results represent a reduction of size and loss by approximately three and one orders of magnitude respectively, and pave the way for high density integration of micron-scale waveguide circuits. The detailed bend design and the experimental work have been carried out on our platform [9], which is based on 4 µm thick single-mode rib waveguides [10, 11] fabricated on silicon-on-insulator (SOI) wafers. However, the same bend concept can be directly applied to any of the 1–10 µm SOI waveguide thicknesses used by us and others, and it can also be adopted on any other HIC waveguide platforms.

**Reducing the bend-size by locally manipulating the waveguide cross-section or the bending radius**

The most efficient way to reduce the minimum bending radius of a waveguide is to increase its lateral index contrast, i.e. the refractive index difference between the waveguide core and the cladding material beside it. This can also be done locally to create short multi-mode waveguide sections within a single-mode waveguide circuit. In past years we have developed a multi-step patterning process [9] that allows us to fabricate not only single-mode rib waveguides, but also through-etched strip waveguides and total internal reflection (TIR) mirrors [12] on SOI wafers (see Figure 1a). The 90º TIR mirrors have an effective bending radius comparable to the waveguide width, and we have repeatedly demonstrated them with ~0.3 dB/90º loss. In comparison, using 4 mm bending radius in our 4 µm thick rib waveguides produces a total bending loss of <0.1 dB/90º, which mainly originates from the same 0.1 dB/cm propagation loss that we measure from our straight waveguides. By construction, the rib waveguide has low lateral index contrast (see Figure 1a), which enables single-mode operation, but also severely limits the minimum bending radius. Both the size of the bent waveguides and the loss of the TIR mirrors significantly limit the dense integration of complicated circuits on the 4 µm SOI platform.

We have also used multi-step patterning for the fabrication of very effective rib-strip converters [13]. This elementary building block is depicted in Figure 1a and it allows passing, without any significant additional loss (<<0.1 dB), from the mode of a single-mode rib waveguide to the fundamental mode of a multi-mode strip waveguide. Unlike the rib waveguide, the strip waveguide in SOI has a very high lateral index contrast, which in principle could allow for almost lossless micron-scale bending radii, just like in the submicron SOI waveguides. But in order to ensure single-mode operation in the lateral direction, the 4 µm thick SOI strip waveguide cannot be wider than 350 nm. Actually, this waveguide width is achievable with our present technology, and we have experimentally shown its feasibility with bending radii as small as 3 µm. Unfortunately, waveguides with such high aspect ratio can suffer from stress-induced cracks or detach from the substrate, which lowers the fabrication yield. Polarization dependency, higher sensitivity to fabrication errors and roughness-induced losses also become major problems with these submicron wide waveguides. Because of all of these reasons, we looked for a solution that could allow for the use of wider (≥1 µm) waveguides and still reach for significantly reduced bending radii.

The only result published in the literature and trying to significantly shrink the bending radius of multi-mode HIC waveguides [14] is based on the so called matched bend approach [15]. It relies on matching the length of an arc (i.e. a circular bend with constant curvature) to an integer multiple of beating lengths between the fundamental mode and the first HOM of the bent waveguide. This approach ensures that light will be efficiently coupled between straight single-mode waveguides, despite the fact that HOMs have been excited in the bent section between them. Anyway, the result presented in [14] still featured a relatively large 195 µm bending radius, which is two orders of magnitude larger than the used 2.95 µm waveguide width.

The size and loss of bends can also be reduced by gradually changing the waveguide cross-section or the bending radius along the bend. This can help to minimize the excitation of HOMs and to avoid the mode field mismatch between straight and bent sections. As the first example, Koos et al. gradually increased the width of a single-mode submicron waveguide along the bend [16]. Anyway this approach is much different from our present solution, is less versatile, and is revealed also to have non-negligible losses.

A well-known approach to reduce both transition and radiation losses in single-mode waveguides is based on bend shapes with continuously varying curvature. Many curve functions have been proposed and explored in the literature [17-20]. However, the previous applications of continuously varying curvature haven't led to any dramatic footprint reduction, but just provided some slight improvement of losses and/or bend size. In references [18] and [19] the same approach has been also applied to some multi-mode medium index contrast waveguides, but attained bend size reductions never reached one order of magnitude, and bend sizes were always two orders of magnitude larger than waveguide widths, mainly because the chosen index contrast was low.

Our earlier approach in applying multi-step patterning to HIC waveguide bends also involved a variable bending radius [20]. We then etched an additional groove at the outer edge of a rib waveguide bend (see Figure 1b) to improve the outermost index contrast of the (4 or 9 µm thick) SOI rib waveguide. The realised test structures demonstrated that the bending radius could be reduced e.g. from 4 to 1 mm without increasing the bending loss. However, despite this and many other significant efforts during the past two decades, no one has been able to demonstrate tight bends in multi-mode optical waveguides or in any micron-scale waveguides with bending radii comparable to the waveguide width. By using strip waveguide bends, like the one depicted in Figure 1c, we were instead able to design, fabricate and measure unprecedented small bends with very low losses on our 4 µm SOI waveguide platform, as will be described in detail below.

## **Design of micron-scale bends in micron-scale waveguides**

By definition modes of a multi-mode straight waveguide propagate without coupling with each other, unless some perturbation occurs, like a change in the waveguide shape. In particular, waveguide bends can induce significant coupling between different modes, so that light is coupled from the fundamental mode in a straight input waveguide to the HOMs of a straight output waveguide placed after the bend. Increasing the curvature $1/R$ (i.e. reducing the bending radius $R$) increases this unwanted coupling and, in general, leads to higher number of significantly excited modes in both the bend and the straight output waveguide. We point out that our following analysis and simulations are performed on the basis of straight waveguide modes, that has been proven to accurately describe also bent multi-mode sections [21,22]. Unless otherwise stated, all simulations presented here assume 1.55 µm wavelength and TE polarization. However, it was confirmed by simulations and experiments that the bends can be designed to work at both polarizations.

In Figure 2a we show simulated power distribution among different modes of a 2 µm wide straight silicon strip waveguide at the end of a 90º arc bend as a function of R. It is clear that light from the fundamental mode in the straight input waveguide is significantly coupled to HOMs whenever small bending radii are used. In these HIC strip waveguides all the power is coupled between propagating modes and coupling to radiation modes is negligible. At R≈11 µm there is first resonant coupling to the fundamental (0th order) mode, but with poor suppression of coupling to 1st, 2nd and 3rd HOMs, resulting in about 90% transmission for the fundamental mode. The first practically useful resonance, i.e. the lowest order low loss matched bend, corresponds to R≈34.4 µm, with fundamental mode transmission > 99%. For larger R values other matched bends occur and, in practice, all HOMs can be neglected but the first one. The power oscillations between this mode and the fundamental mode slowly damp with R, and for R>400 µm maximum coupling to the HOM is suppressed by more than 20 dB. We will adopt this suppression level as the threshold to define the minimum R value ensuring low-loss non-resonant operation of the bend. This corresponds to 1% loss (0.044 dB) in the fundamental mode. Unlike the matched bend case, where power is significantly coupled to HOMs in the bent section and then completely coupled back to the fundamental mode at the very end of the bend, proper non-matched operation requires coupling to HOMs to be always suitably suppressed during propagation. In other words, the matched-bend is a resonant system, whereas the generic non-matched bend is adiabatic. It is clear that non-matched operation ensures broader operation bandwidth and higher tolerance to fabrication errors. In general, in any bend of any shape we can distinguish between two working principles. In resonant operation the bend length is matched to the beating length between the fundamental mode and the first HOM - ensuring high coupling back to the fundamental mode at the very end of the bend. In adiabatic operation one simply ensures low coupling to HOMs at any propagation step. We will refer in the first case to a "matched bend" and in the second case to an "adiabatic bend".

The idea behind our design approach was to use so-called Euler bends that have a unique bend shape. To bend light by any given angle $\theta$ we use an Euler bend that consists of two mirror-symmetric sections, each of them bending light by $\theta/2$. The curvature is changed linearly with the bend length *s* as shown in Figure 3a. For example, Figure 3b and Figure 3c show an Euler L-bend and an Euler U-bend with normalised minimum radius $R_{min}$. We define the effective bending radius $R_{eff}$ to be the radius of the equivalent arc that would join the starting and ending points of the Euler bend. For the Euler L-bend $R_{eff}=1.87\ R_{min}$, and for the Euler U-bend $R_{eff}=1.38\ R_{min}$. The same approach can be also used to design S-bends, through a doubly symmetric structure.

The mathematical function describing a bend with the curvature varying linearly along the path length is the so called Euler spiral [23] and this is why we refer to our mirror-symmetric bends as Euler bends. This curve is also known as Cornu spiral or clothoid, and it is widely used as a track transition curve in civil engineering for road and railway construction [24], but it has had limited use in guided-wave optics [25,26]. The only prior example available in the literature of using this bend shape in a strongly-confined multi-mode waveguide doesn't pertain to the optical domain. In fact, it is a very old application of the Euler spiral to microwaves propagating in a dielectric coated metal waveguide with circular cross-section [27], and was mainly meant to suppress coupling to the degenerate orthogonal polarization mode.

For the first time, we applied the Euler bend geometry to HIC multi-mode strip waveguides and found up to 3 orders of magnitude improvement in effective bending radius compared to arc bends. The Euler bends were also found to have smaller size, wider bandwidth and more relaxed tolerances to fabrication errors when compared to matched arc bends. For example, Figure 2b shows power coupling to different modes at the output of a 2 µm wide Euler L-bend as a function of the effective radius. The adiabatic Euler bend has $R_{eff}$=75 µm, being more than 5 times smaller than the adiabatic arc. The corresponding minimum radius is 40 µm, i.e. 1.87 times smaller than $R_{eff}$, as stated above. Furthermore the first useful matched bend occurs at $R_{eff}$=16.6 µm, i.e. at less than half the size of the smallest matched arc. The second one, at $R_{eff}$=37.4 µm, is comparable in size with the arc bend, but ensures much better performance. In fact it is interesting to analyse the spectral response of the bends, which is also a measure of their tolerance to fabrication errors, because what matters in practice is the ratio between the waveguide size and the wavelength. In Figure 4 we compare the spectral responses of different bends. Notice that in all cases the matched bends are not exactly set to the transmission peak for 1.55 µm wavelength, but are optimized a little bit off resonance to ensure the highest operation bandwidth. Besides size reduction, a comparison between the smallest matched arc (Figure 4a, R=33.5 µm) and the smallest matched Euler L-bend (Figure 4c, R=17.2 µm) highlights an order of magnitude broader bandwidth for the Euler L-bend. In the most interesting wavelength range at 1.2–1.7 µm the second matched Euler L-bend (Figure 4d, R=39.8 µm) performs better than the smallest adiabatic arc (Figure 4b, R=400 µm), and even better than the adiabatic Euler L-bend (Figure 4e, R=74.8 µm). These simple examples show that matched and adiabatic Euler bends can be not only much smaller than corresponding matched and adiabatic arc bends, but they also perform much better in terms of bandwidth and tolerances to fabrication errors. Similar results hold for the other waveguide widths and bend angles that we simulated. One example is the ultra-compact Euler U-bend shown in Figure 3d, which we discuss more below.

## **Experimental demonstration of the miniature bends on 4 µm SOI platform**

Cascades of 4 to 44 bends were designed and fabricated on our mature SOI platform to experimentally measure the bending losses of 1 and 2 µm wide waveguides. Most of the bends had losses below the maximum accuracy of our measurement system, which is about ±0.02 dB per bend. For example, all S-bends (20°, 9° or 4° angle) featured negligible losses, as well as all U- and L-bends with larger bending radii ($R_{eff}$>20 µm). In Figure 5a we show some results with non-negligible losses. An ultra-small U-bend, like the one simulated in Figure 3d, with $R_{eff}$=1.27 µm and 0.9 µm waveguide width featured 0.09 dB and 0.17 dB per bend loss for TE and TM polarization, respectively. We point out that this is the smallest bend of a semiconductor waveguide reported to date, and also has significantly lower loss compared to previous achievements [16].

U-bends with $R_{eff}$=5.4 µm (1 µm width) and $R_{eff}$=28.8 µm (2 µm width), as well as L-bends with effective radii of 16.8 µm and 39 µm (2 µm width) all showed 0.02–0.03 dB losses per bend at both polarizations, which is at the limit of our measurement accuracy. The spectral response of cascaded L-bends in a 1 µm wide waveguide with $R_{eff}$=13.1 µm is compared in Figure 5b with the one of a cascade of 44 TIR mirrors. This clearly shows the negligible loss per bend and the broadband spectral operation of the designed bends. Most of the polarisation dependent losses were measured to be within our measurement accuracy.

# **Conclusions**

In summary, we have demonstrated with both simulations and experiments that ultra-small waveguide bends with a micron-scale core, high index contrast and Euler spiral shape can have very low losses. Thanks to higher field confinement, the performance of these bends reaches beyond submicron waveguides, ensuring at the same time lower radiation losses and reduced impact of side-wall roughness [16],[26]. Furthermore they were found to be tolerant to fabrication errors and to work properly on wide wavelength ranges. The bends in micron-scale multi-mode strip waveguides are readily interfaced to single-mode rib waveguides with mode size comparable to optical fibres. By completely removing the main drawback of the micron-scale waveguide circuits - namely the large bending radii required to ensure small bending losses - we expect the present breakthrough to lead to dramatic shrinking of photonic device size not only in our silicon platform but also in all other commercial semiconductor platforms, such as InP and GaAs, leading to more advanced circuits, significant cost reductions and extensive market penetrations. The device footprints can now be made comparable to those on submicron waveguide platforms, while avoiding many critical issues that are presently faced in dense photonics integration, such as high sensitivity to fabrication errors, polarization dependence, difficult fibre coupling and high cost of fabrication tools.

The new Euler bend concept can also be used to enable new functionalities in micron-scale waveguide platforms, such as compact multi-mode interferometric reflectors [14], densely integrated cross-connectors and clock-distribution circuits, very long spirals within a very small footprint (e.g. 10 cm waveguide in a 0.3 mm$^2$ area), compact arrayed waveguide gratings, small and low loss Mach-Zehnder interferometric filters, and high-finesse microring resonators.

## Methods

### *Design and simulations*

Light propagation in the bends was simulated using FIMMPROP software. An Euler bend with an angle θ is simulated as a cascade of 60 arc sections with constant θ/60 bending angle and varying curvature 1/$R$. The length of each section is easily determined from the curvature of the Euler bend, i.e. $1/R = d\theta/ds \propto \theta^2$, where *s* is the path length. Each arc section is treated as a cascade of straight sub-sections bent with respect to each other by a constant angle. The modes of the straight waveguide are calculated by a fully vectorial mode solver taking into account the wavelength dependence of the refractive indices, and the scattering matrix between modes is evaluated in each arc section (regarded as a periodic system), to give the overall scattering matrix of the whole bend. We found perfect agreement between 3D simulations (i.e. assuming the real 2D waveguide cross section) and much faster 2D simulations (using the effective index method to shrink the waveguide in the vertical direction). Furthermore, the found resonance values of the Euler matched-bend effective radii have been confirmed by both 2D Finite Difference Time Domain and 2D Finite Element Frequency Domain simulations of the same bend shapes (like the one shown in Figure 3.d). For design purposes, the Euler bend can be easily calculated through expansion series of Fresnel integrals [23, 28], and for our purposes already 2 to 3 expansion terms give reasonable accuracy.

### *Device fabrication*

The devices were fabricated using smart-cut silicon-on-insulator wafers from SOITEC. Initial SOI layer thickness was increased with epitaxial silicon growth. FilmTek 4000 spectrophotometer was used to measure the thicknesses of the SOI layer (4.20 ± 0.04 µm) and the underlying buried oxide (2.998 ± 0.002 µm). A 0.5 µm thick TEOS oxide layer was deposited on the wafer in LPCVD diffusion furnace to work as a hard mask in silicon etching. Waveguide fabrication was done using our standard double-masking multi-step process [9]. In the process, two mask layers are passively aligned with respect to each other, and two separate silicon etch-steps form the rib and strip waveguide structures and waveguide facets. Lithography steps were done using FPA-2500i3 i-line wafer stepper from Canon Inc. and pattern transfer to oxide hard mask was done using LAM 4520 reactive ion etcher with $CF_4$ and $CHF_3$ chemistry. Waveguides were etched into silicon using Aviza ICP etcher from SPTS Technologies. This was done with a modified Bosch process [29] using $SF_6$ and $C_4F_8$ as etch and passivation gases, respectively, and $O_2$ as an etch gas to break passivation polymer formed by $C_4F_8$. After silicon etching, TEOS hard mask was removed with buffered oxide etch (BOE). Hard mask removal was followed by wet thermal oxidation of 0.5 µm, and thermal oxide etching with BOE. This was done to smoothen the etched surfaces, and to thin the SOI-layer to its final thickness of approximately 4 µm. A 0.17 µm thick silicon nitride layer was then deposited on the wafer with LPCVD, as an antireflection coating and to prevent the etching of buried oxide in later oxide wet etch process steps. This layer was patterned in hot phosphoric acid using a hard mask made of 0.5 µm thick LPCVD TEOS and patterned with BOE. Another 0.5 µm thick LPCVD TEOS layer was then deposited to form a cladding oxide, and locally removed from waveguide facets with BOE before dicing the wafer into chips. Whenever damage to the end-facets was detected after dicing, these facets were polished to optical quality using diamond lapping films.

*Optical measurements*

In the loss measurements, light at 1.55 nm wavelength from a Fabry-Perot laser (HP 81554SM) was coupled from a polarization-maintaining fiber (Flextronics) into a waveguide via a polarization controller. All measurements were done separately for TE and TM polarization. The transmitted light from the output waveguide was coupled into an optical power meter via single-mode fiber. The use of both single-mode fiber and input/output single-mode rib waveguides on chip guaranteed that measurements were done for the fundamental mode only. Index matching oil (refractive index 1.5) was used to reduce reflections at the polished waveguide facets. The width of each waveguide was increased to 10 µm at the chip facet to improve fiber coupling efficiency. In the spectral measurements, a superluminescent light emitting diode was used as a broadband light source. Output spectrum with range from 1400 nm to 1700 nm was measured with an optical spectrum analyser (ANDO AQ-6315A). The reported transmission values include losses from coupling to input and output fibers, which are not optimized to the used waveguide dimensions.

**Figure legends**

Figure 1. **Micron-scale silicon photonics platform. a.** Single mode rib waveguides can be tightly bent by TIR mirrors with 0.3 dB/90⁰ loss; they can be also be turned into strip waveguides by almost lossless converters; **b.** groove bends have been previously explored to reduce the bend size without affecting the losses; **c:** in this work we propose to use suitably designed bends of multimode strip waveguides to dramatically reduce bend size and losses.

Figure 2. **Simulated TE polarization coupling to higher order modes at the output of multimode 90⁰ bends as a function of the radius**. **a.** Arc in a 2 µm wide and 4 µm thick silicon strip waveguide. **b.** Euler L-bend with the same waveguide structure.

Figure 3. **Euler bend layouts**. **a.** Linear change of curvature as a function of path length $s$ in a mirror-symmetric Euler-bend. **b.** Example application to a 90⁰ bend (L-bend) with normalized minimum bending radius $R_{min}=1$. The corresponding effective bending radius $R_{eff}$ is highlighted. **c.** Same as b. but for a 180⁰ bend (U-bend); **d.** 2D Finite Element Frequency Domain simulation of light propagation in a U-bend with 900 nm waveguide width and 960 nm minimum bending radius (electric field, TE polarization). The field pattern clearly shows how, in a matched-bend, HOMs are indeed excited, but finally coupled back to the fundamental mode.

Figure 4. **Simulated excitation of TE modes at the output of different bends as a function of wavelength**. All bends are made in 2 µm wide and 4 µm thick silicon strip waveguides. **a.** Matched 90⁰ arc with 33.5 µm bending radius. The rectangle highlights the narrow operation bandwidth; **b.** Adiabatic arc with 400 µm bending radius; **c**. Euler L-bend with an effective radius of 17.2 µm. Its wide operation bandwidth is highlighted; **d**. Matched Euler L-bend

with an effective radius of 39.8 μm; **e**. Adiabatic Euler L-bend with an effective radius of 74.8 μm.

Figure 5. **Experimental results. a**. Measured scaling of losses with the number of bends for different U-bends (at 1.55 μm wavelength); **b.** Spectral responses of cascaded Euler L-bends and 44 TIR mirrors (TE polarization); **c.** SEM picture of some waveguide structures patterned into a 4 μm thick SOI layer with different number of bends and different bending radii; **d.** SEM picture of a cascade of four U-bends in a 2 μm wide waveguide with 7.5 μm effective bending radius.

## Acknowledgements

This work was supported by the European Community's Seventh Framework Programme (FP7-ICT/2007-2013) within the RAMPLAS project (ICT- FET 270773).

## Author Contributions

MC designed the bends, MC and MH made the simulations, SY fabricated the structures, which were then characterized by MK. All authors discussed the results and contributed to manuscript preparation.

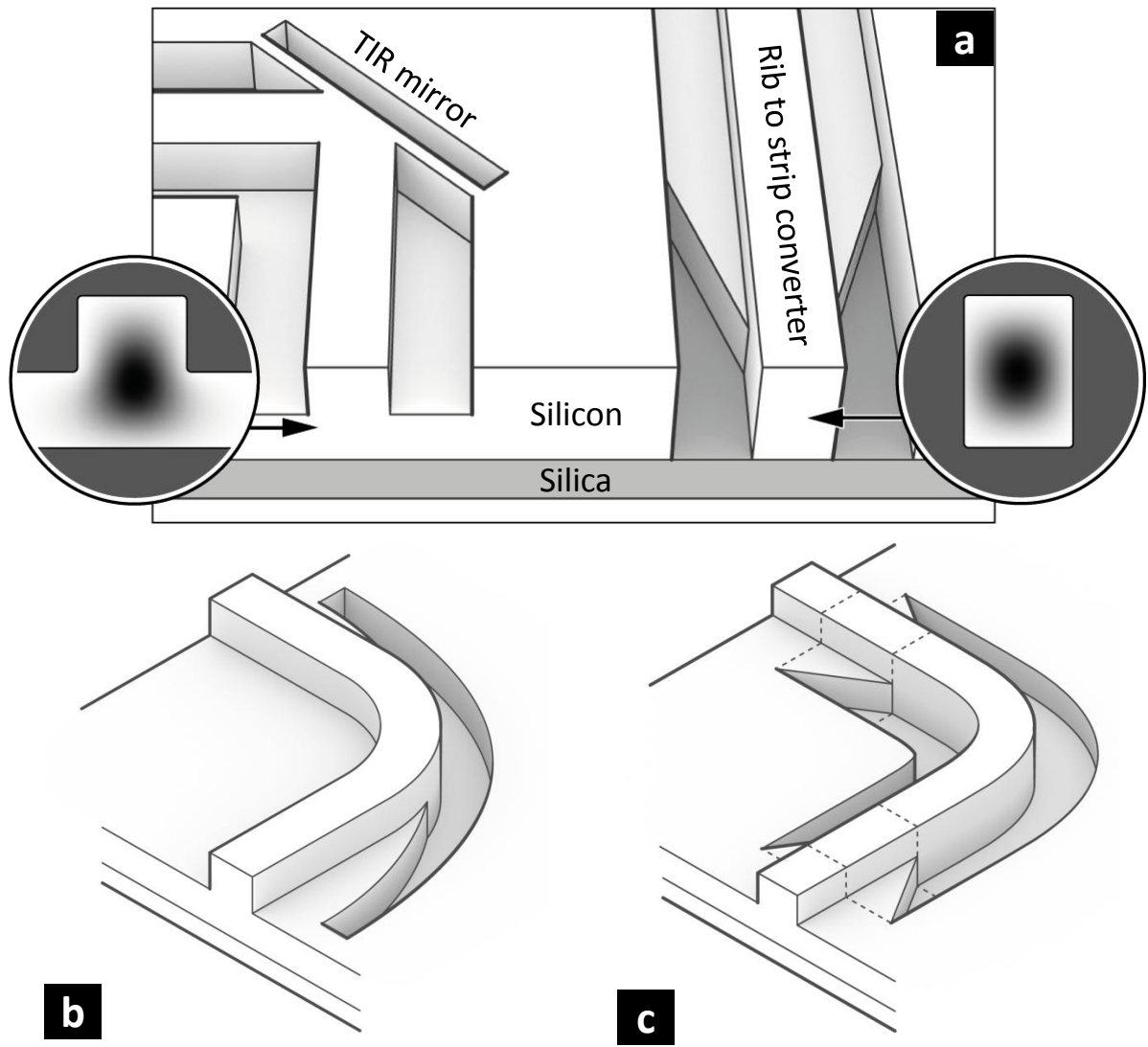

Figure 1

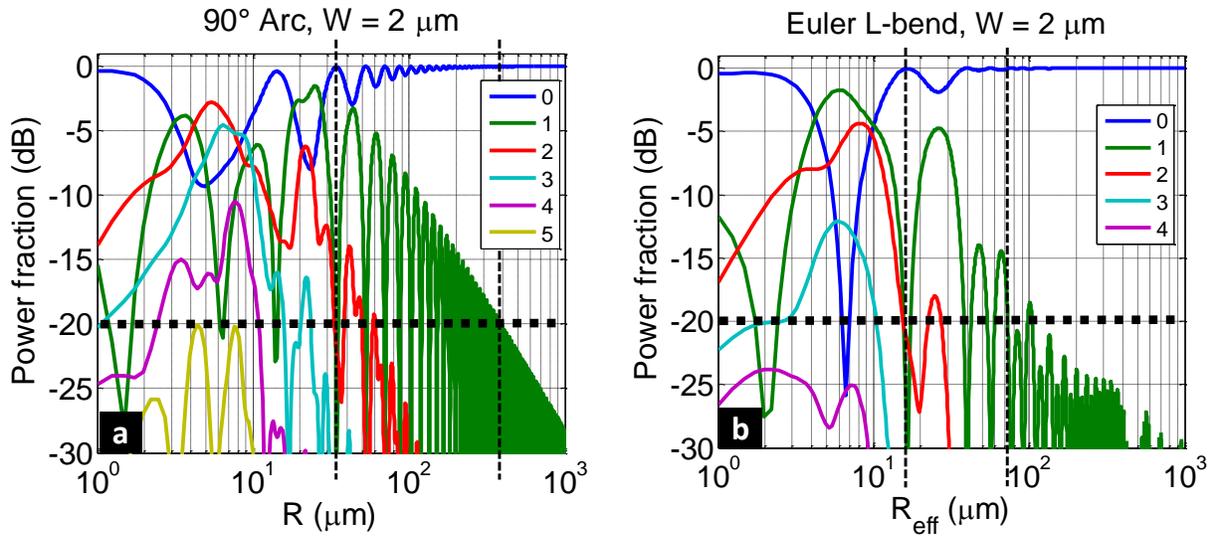

Figure 2

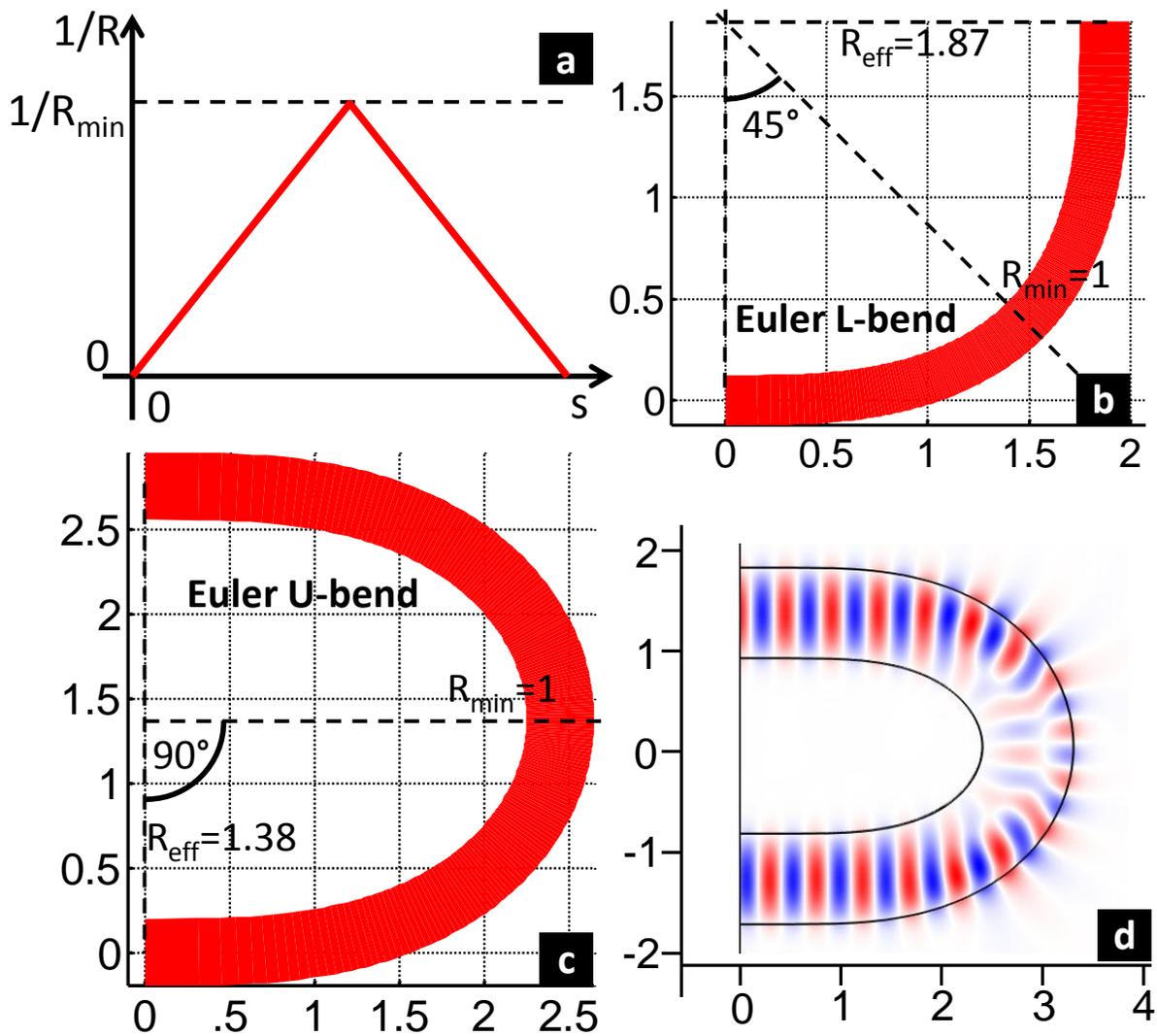

Figure 3

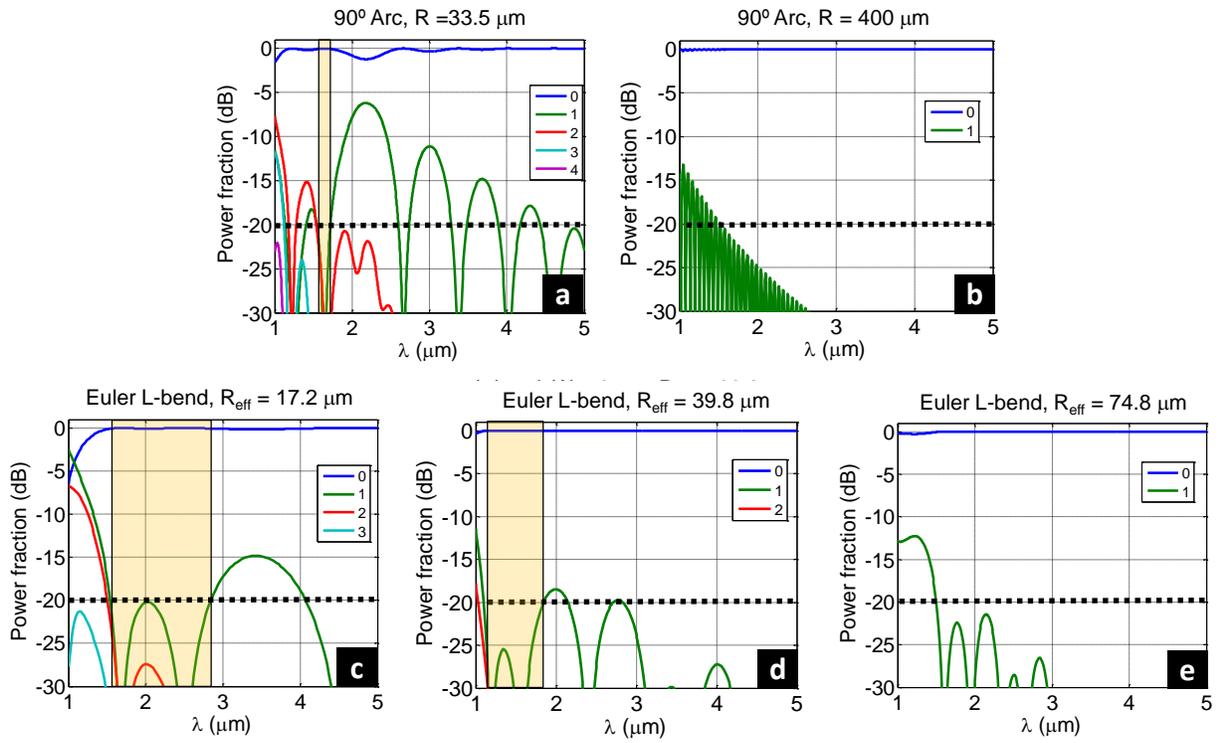

Figure 4

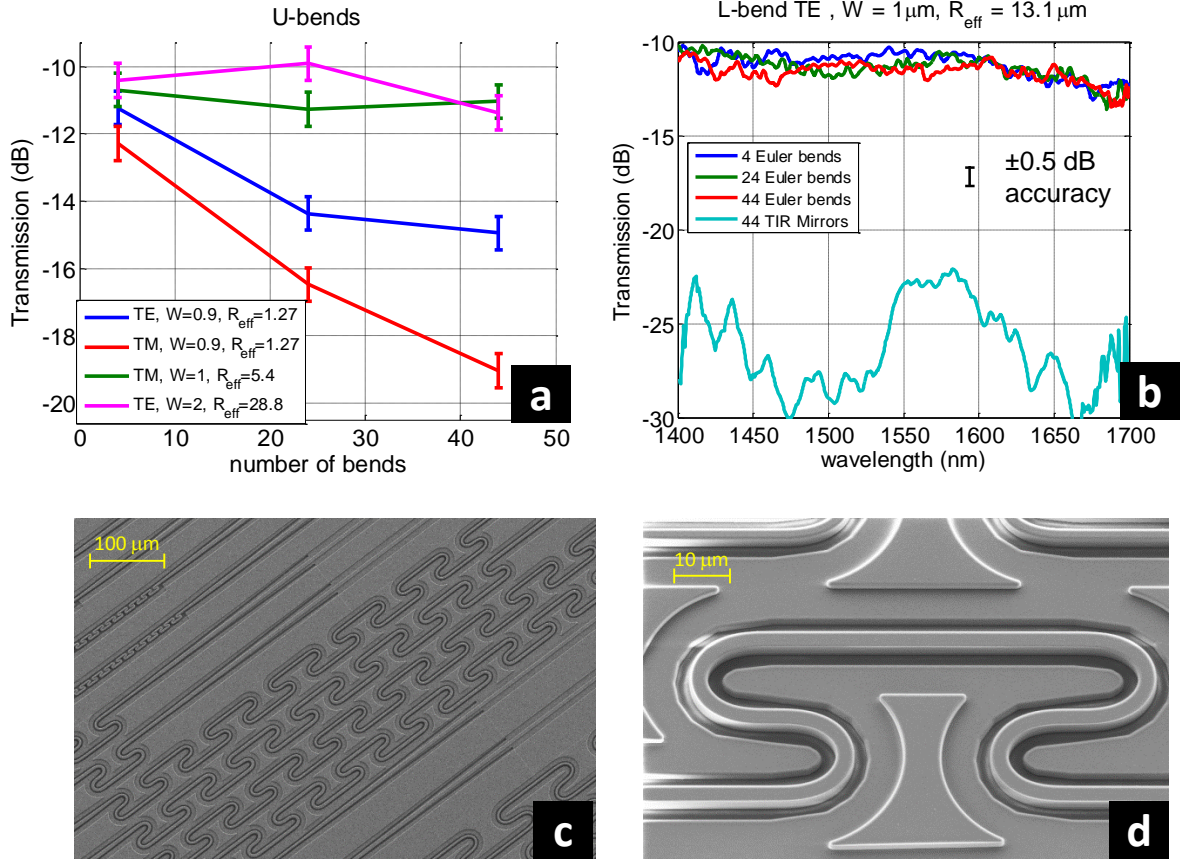

Figure 5